# EIGENRAYS IN 3D HETEROGENEOUS ANISOTROPIC MEDIA:
# PART VI – DYNAMICS, LAGRANGIAN VS. HAMILTONIAN APPROACHES

*Igor Ravve (corresponding author) and Zvi Koren, Emerson*

*igor.ravve@emerson.com , zvi.koren@emerson.com*

## ABSTRACT

In Part V of this study, we presented an original Lagrangian approach for computing the dynamic characteristics along stationary rays, by solving the linear, second-order Jacobi differential equation, considering four sets of initial conditions as the basic solutions (two for point-source and two for plane-wave). We then focused on the computation of the Green function amplitude (the geometric spreading) and phase changes due to caustics, where only the two point-source basic solutions with their corresponding initial conditions are required. Solutions of the Jacobi equation represent the normal shift vectors of the paraxial rays and define the geometry of the ray tube with respect to the stationary central ray.

Rather than the Lagrangian approach, the dynamic characteristics are traditionally computed with the Hamiltonian approach, formulated normally in terms of two first-order differential equations, where the solution variables are the paraxial shifts and paraxial slowness changes along the ray. In this part (Part VI), we compare and relate the two approaches. We first combine the two first-order Hamiltonian dynamic equations, eliminating the paraxial variations of the slowness vector. This leads to a second-order differential equation in terms of the Hamiltonian shift alone, whose ray-normal counterpart coincides with the normal shift of the corresponding Lagrangian solution, while the ray-tangent component does not affect the Jacobian and the geometric spreading. Comparing the proposed Lagrangian approach to the dynamic ray tracing with the "classical" Hamiltonian approach, we demonstrate that they are fully compatible/equivalent for general anisotropy. We then derive the two-way relationships between the Hamiltonian's and Lagrangian's Hessians, which are the core computational elements of dynamic ray theory.

Finally, we demonstrate the relationships between these two types of the Hessians analytically, for an example based on an ellipsoidal orthorhombic medium, and numerically, for a general triclinic medium.

## INTRODUCTION

This part of our study is a direct continuation of Part V. We elaborate on the relations between the proposed Lagrangian approach and the commonly used Hamiltonian approaches, demonstrating their theoretical equivalences and comment on the main differences, and emphasize the advantages of our proposed Lagrangian approach. The readers who like to proceed to the finite-element implementation of the theory presented in Part V, may omit this part, and move to Part VII.

Applying the proposed arclength-related Lagrangian $L(\mathbf{x},\mathbf{r})$ and its matching Hamiltonian $H(\mathbf{x},\mathbf{p})$, we derive the two-way relationships between all their corresponding Hessian matrices. In particular, we relate the Hessian tensors $L_{\mathbf{rr}}$ and $H_{\mathbf{pp}}^{-1}$, which are the core computational elements in dynamic ray theory. While both of them represent the gradient of the slowness vector $\mathbf{p}$, with respect to (wrt) the ray direction vector $\mathbf{r}$, $\partial \mathbf{p}/\partial \mathbf{r}$, these are different tensors. The directional Hessian of the Lagrangian, $L_{\mathbf{rr}}$, is a singular matrix, with a vanishing determinant, while the slowness Hessian of the Hamiltonian, $H_{\mathbf{pp}}$ and its inverse, $H_{\mathbf{pp}}^{-1}$, are regular invertible matrices, with the exception of inflection points along the ray path (Bona and Slawinski, 2003). We interpret the physical nature of this discrepancy between the two matrices, and then demonstrate that the eigensystems of these matrices are fairly close, with a single divergent parameter.

The connection between the Hamiltonian's and Lagrangian's Hessians is then illustrated using two numerical examples: analytically, for an ellipsoidal orthorhombic medium, and numerically, for a triclinic medium.

Appendices

In order to make the paper more readable, the body of the paper contains the main theoretical concepts with the principal governing equations, with minimum mathematical derivations. The detailed derivations have been moved to the appendices.

In Appendices A and B, we demonstrate that both Cartesian dynamic ray tracing (DRT) formulations: the proposed Jacobi Lagrangian-based second-order ODE, and the commonly used Hamiltonian-based first-order ODE set (e.g., Červený, 2000), are fully consistent for isotropic media (Appendix A) and for general anisotropic media (Appendix B). We then derive the relationships between the second derivatives of the Lagrangian wrt the location and direction vector components, and those of the Hamiltonian wrt the location and slowness vector components, for isotropic and general anisotropic media.

In Appendix C, we demonstrate the relationships presented in Appendix B numerically, using a spatially varying triclinic medium, and analytically, for an ellipsoidal orthorhombic medium.

In appendix D, we further explain the physical meaning of the (ray-direction based) plane-wave initial conditions (IC) used in this study. Our definition of the paraxial plane wave differs from the conventional, standard one: We assume that the direction of the paraxial plane wave at the source point is collinear with that of the central ray (ray direction), while normally in the literature, the source plane waves are defined such that their paraxial slowness is collinear with the slowness vector of the central ray. To explore this discrepancy, we study the difference between the paraxial and central slowness directions for a plane-wave at the source.

## HAMILTONIAN AND LAGRANGIAN APPROACHES TO DYNAMIC RAY TRACING

Both, kinematic and dynamic ray tracing (KRT and DRT) can be performed either with the Lagrangian or with the Hamiltonian approaches, depending on the problem to be solved. In this section, we compare our proposed Lagrangian-based approach with the alternative Hamiltonian-based approach, demonstrating their theoretical equivalence, and provide the two-way relationships between the Hamiltonian's and Lagrangian's Hessians.

Lagrangian approach

As demonstrated in Part V, the Lagrangian dynamics workflow for computing the geometrical spreading includes three stages:

a) Obtaining the normal paraxial shifts vectors,

$$\mathbf{u}_i(s) = \partial \mathbf{x}_{\text{prx}} / \partial \gamma_i , \quad i = 1, 2 \quad , \tag{1}$$

where $\gamma_i$ are the ray coordinates (RC). For this, we solve the Jacobi equation twice,

$$\frac{d}{ds}\left(L_{\mathbf{rx}}\cdot\mathbf{u}+L_{\mathbf{rr}}\cdot\dot{\mathbf{u}}\right)=L_{\mathbf{xx}}\cdot\mathbf{u}+L_{\mathbf{xr}}\cdot\dot{\mathbf{u}} \quad , \tag{2}$$

applying two different point-source initial conditions, respectively,

$$\begin{array}{llll} L_{\mathbf{rr},S}\dot{\mathbf{u}}_{1,S}=\lambda_{1,S}\dot{\mathbf{u}}_{1,S} \;, & L_{\mathbf{rr},S}\dot{\mathbf{u}}_{2,S}=\lambda_{2,S}\dot{\mathbf{u}}_{2,S} \;, & L_{\mathbf{rr},S}\mathbf{r}_S=0 \;, \\ \mathbf{u}_{1,S}=0 \;, & \mathbf{u}_{2,S}=0 \;, & \dot{\mathbf{u}}_{1,S}\times\dot{\mathbf{u}}_{2,S}\cdot\mathbf{r}_S=1 \end{array} \tag{3}$$

where he subscript $S$ refers to the source point, $\dot{\mathbf{u}}_{1,S}$ and $\dot{\mathbf{u}}_{2,S}$ are the normalized eigenvectors of the directional Hessian matrix at the source, $L_{\mathbf{rr},S}$, while $\lambda_{1,S}$ and $\lambda_{2,S}$ are their corresponding nonzero eigenvalues.

b) Computing the ray Jacobian at any point along the ray,

$$J(s)=\mathbf{u}_1(s)\times\mathbf{u}_2(s)\cdot\mathbf{r}(s) \quad , \tag{4}$$

and eventually,

c) Establishing the relative geometric spreading,

$$L_{GS}(s)=\sqrt{\frac{v_{\mathrm{ray},S}}{v_{\mathrm{phs},S}}\frac{v_{\mathrm{ray}}(s)}{v_{\mathrm{phs}}(s)}\frac{|J(s)|}{\lambda_{1,S}\lambda_{2,S}}} \quad . \tag{5}$$

Hamiltonian approach

The Hamiltonian DRT set is usually formulated as a system of two first-order ordinary differential equations (e.g., Červený, 2000), involving the paraxial Hamiltonian shifts, $\mathbf{w}$, and the paraxial slowness variations, $\mathbf{p}_{\text{prx},\gamma}$,

$$\dot{\mathbf{w}} = H_{\mathbf{px}}\mathbf{w} + H_{\mathbf{pp}}\mathbf{p}_{\text{prx},\gamma} \quad , \quad \dot{\mathbf{p}}_{\text{prx},\gamma} = -H_{\mathbf{xx}}\mathbf{w} - H_{\mathbf{xp}}\mathbf{p}_{\text{prx},\gamma} \quad . \tag{6}$$

Conventional derivations of the KRT and DRT equations are based on the Hamiltonian conservation principle, where, the Hamiltonian remains constant along the ray. For the DRT, the Hamiltonian also remains constant for any varying ray coordinate (RC), and this leads to a scalar constraint accompanying the conventional DRT equation set (e.g., Červený, 2000, equation 4.2.7). However, the formula for the constraint is not unique, and it depends on the form of the Hamiltonian. Considering different forms of the Hamiltonian (Appendix A for isotropic media and Appendix B for general anisotropic media), we demonstrate that the proposed linear, second-order Jacobi DRT equation is consistent with different forms of the linear, first-order Hamiltonian DRT equation set, provided the proper constraint, corresponding to the chosen flow parameter (such as traveltime, arclength or sigma), is imposed. Moreover, we also show in Appendix A that different forms of constraint may correspond even to the same flow parameter – the arclength. A general form of the constraint, accompanying the DRT equation set 6, reads,

$$\frac{dH}{d\gamma} = H_{\mathbf{x}}\mathbf{w} + H_{\mathbf{p}}\mathbf{p}_{\text{prx},\gamma} = 0 \quad . \tag{7}$$

where the Hamiltonian-based paraxial shifts and the slowness variations are defined by,

$$\mathbf{w}_i = \partial \mathbf{x}_{\text{prx}}^H / \partial \gamma_i \quad , \quad \mathbf{p}_{\text{prx},\gamma} = \partial \mathbf{p}_{\text{prx}} / \partial \gamma_i \quad . \tag{8}$$

The superscript $H$ emphasizes that the paraxial ray location $\mathbf{x}_{\text{prx}}^H$ has been obtained with the Hamiltonian approach; it differs from the Lagrangian $\mathbf{x}_{\text{prx}}$, but, as explained below, this difference is inessential,

$$\mathbf{x}_{\text{prx}} = \sum_{i=1}^{4} \gamma_i \mathbf{u}_i \quad , \quad \mathbf{x}_{\text{prx}}^H = \sum_{i=1}^{4} \gamma_i \mathbf{w}_i \quad , \quad (9)$$

where the RC, $\gamma_i$, $i = 1,2\ldots4$, for the Lagrangian-based and Hamiltonian-based descriptions of paraxial rays are identical. The Lagrangian shift $\mathbf{u}_i$ represents the ray-normal vector, while the Hamiltonian shift $\mathbf{w}_i$ consists of both ray-tangent and ray-normal components,

$$\mathbf{w}(s) = \mathbf{u}(s) + \rho_t(s)\mathbf{r}(s) \quad , \quad (10)$$

where $\rho_t(s)$ is a scalar function.

The Lagrangian and Hamiltonian solutions, $\mathbf{u}$ and $\mathbf{w}$, respectively, differ only by this inessential tangent counterpart, $\rho_t \mathbf{r}$ (see also Appendix B), which has no effect on the ray Jacobian (and thus, neither on the relative geometric spreading),

$$\mathbf{w}_1 \times \mathbf{w}_2 \cdot \mathbf{r} = \left(\mathbf{u}_1 + \rho_{t,1}\mathbf{r}\right) \times \left(\mathbf{u}_2 + \rho_{t,2}\mathbf{r}\right) \cdot \mathbf{r} = \mathbf{u}_1 \times \mathbf{u}_2 \cdot \mathbf{r} = J \quad . \quad (11)$$

In Appendix B we show that for the arclength-related Hamiltonian, the general constraint of equation 7 simplifies to,

$$\dot{\mathbf{w}}(s) \cdot \mathbf{r}(s) = 0 \quad , \quad (12)$$

which resolves the tangent counterpart of the Hamiltonian-based paraxial shift,

$$\dot{\rho}_t(s) = \mathbf{u}(s) \cdot \dot{\mathbf{r}}(s) \quad , \tag{13}$$

where $\dot{\mathbf{r}}$ is the curvature vector of the ray at $\mathbf{x}$. At the start point of the point-source paraxial ray, the tangent counterpart vanishes (as well as the normal one), and it follows from equations 3 and 13, that its derivative also vanishes,

$$\rho_{t,S} = 0 \quad , \quad \dot{\rho}_{t,S} = 0 \quad . \tag{14}$$

This, in turn, means that for point-source rays, the initial conditions for the Lagrangian and Hamiltonian shift vectors are identical,

$$\begin{aligned} \mathbf{w}_{1,S} &= \mathbf{u}_{1,S} \, , \quad \dot{\mathbf{w}}_{1,S} = \dot{\mathbf{u}}_{1,S} \, , \\ \mathbf{w}_{2,S} &= \mathbf{u}_{2,S} \, , \quad \dot{\mathbf{w}}_{2,S} = \dot{\mathbf{u}}_{2,S} \, . \end{aligned} \tag{15}$$

We combine the two first-order equations of set 6, eliminate the slowness variation $\mathbf{p}_{\mathrm{prx},\gamma}$, and apply the constraint of equation 12. This leads to the second-order Hamiltonian DRT equation,

$$\frac{d}{ds}\left(H_{\mathbf{pp}}^{-1}\dot{\mathbf{w}} - H_{\mathbf{pp}}^{-1}H_{\mathbf{px}}\mathbf{w}\right) = -H_{\mathbf{xp}}H_{\mathbf{pp}}^{-1}\dot{\mathbf{w}} + \left(H_{\mathbf{xp}}H_{\mathbf{pp}}^{-1}H_{\mathbf{px}} - H_{\mathbf{xx}}\right)\mathbf{w} \quad . \tag{16}$$

Relationship between the two approaches

Note that due to the first-order homogeneity of the proposed Lagrangian wrt the ray direction vector $\mathbf{r}$, any arclength-dependent vector tangent to the ray is a solution of the Jacobi equation (Bliss, 1916). This means that if the Lagrangian-based shift $\mathbf{u}$ is a solution of equation 2, then the Hamiltonian-based shift $\mathbf{w}$ is its solution as well,

$$\frac{d}{ds}(L_{\mathbf{rx}} \cdot \mathbf{w} + L_{\mathbf{rr}} \cdot \dot{\mathbf{w}}) = L_{\mathbf{xx}} \cdot \mathbf{w} + L_{\mathbf{xr}} \cdot \dot{\mathbf{w}} \quad . \tag{17}$$

Comparing equations 16 and 17, we obtain the two-way relationships between the Hamiltonian's and Lagrangian's Hessians,

$$\begin{aligned} L_{\mathbf{rr}} &= H_{\mathbf{pp}}^{-1} - \lambda_{\mathbf{r}} \mathbf{r} \otimes \mathbf{r} \quad, \quad L_{\mathbf{rx}} = -H_{\mathbf{pp}}^{-1} H_{\mathbf{px}} \quad, \\ L_{\mathbf{xr}} &= -H_{\mathbf{xp}} H_{\mathbf{pp}}^{-1} \quad, \quad L_{\mathbf{xx}} = H_{\mathbf{xp}} H_{\mathbf{pp}}^{-1} H_{\mathbf{px}} - H_{\mathbf{xx}} \quad, \end{aligned} \tag{18}$$

where $\lambda_{\mathbf{r}}$ is the eigenvalue of the inverse Hamiltonian's Hessian wrt the slowness vector $\mathbf{p}$, $H_{\mathbf{pp}}^{-1}$, corresponding to its eigenvector $\mathbf{r}$, and hence,

$$\begin{aligned} H_{\mathbf{pp}}^{-1} &= L_{\mathbf{rr}} + \lambda_{\mathbf{r}} \mathbf{r} \otimes \mathbf{r} \quad, \quad H_{\mathbf{px}} = -H_{\mathbf{pp}} L_{\mathbf{rx}} \quad, \\ H_{\mathbf{xp}} &= -L_{\mathbf{xr}} H_{\mathbf{pp}} \quad, \quad H_{\mathbf{xx}} = H_{\mathbf{xp}} H_{\mathbf{pp}}^{-1} H_{\mathbf{px}} - L_{\mathbf{xx}} \quad . \end{aligned} \tag{19}$$

The detailed derivation is provided in Appendix B.

**DIRECTIONAL GRADIENT OF SLOWNESS: HAMILTONIAN AND LAGRANGIAN**

In this section we define the slowness vector gradient wrt the ray direction vector $\mathbf{r}$, $\mathbf{p_r} \equiv \partial \mathbf{p} / \partial \mathbf{r}$, applying the Lagrangian and the Hamiltonian approaches, where the results prove to be different. We will explain the physical nature of this difference and provide the constraint relating the two matrices.

Lagrangian approach

According to the momentum equation,

$$\mathbf{p} = L_{\mathbf{r}}(\mathbf{x},\mathbf{r}) \quad , \qquad \mathbf{p_r} \equiv \frac{\partial \mathbf{p}}{\partial \mathbf{r}} = L_{\mathbf{rr}}(\mathbf{x},\mathbf{r}) \quad , \qquad (20)$$

where, for the first-degree homogeneous Lagrangian wrt $\mathbf{r}$, the directional Hessian $L_{\mathbf{rr}}$ (which represents also the directional slowness gradient) is a symmetric singular matrix, with a zero eigenvalue corresponding to the eigenvector $\mathbf{r}$.

Hamiltonian approach

With the arclength-related Hamiltonian, we obtain,

$$\mathbf{r} = H_{\mathbf{p}}(\mathbf{x},\mathbf{p}) \quad , \qquad \mathbf{r_p} \equiv \frac{\partial \mathbf{r}}{\partial \mathbf{p}} = H_{\mathbf{pp}}(\mathbf{x},\mathbf{p}) \quad , \qquad (21)$$

which leads to,

$$\mathbf{p_r}^H = \left(\frac{\partial \mathbf{p}}{\partial \mathbf{r}}\right)_H = \left(\frac{\partial \mathbf{r}}{\partial \mathbf{p}}\right)_H^{-1} = H_{\mathbf{pp}}^{-1}(\mathbf{x},\mathbf{p}) \quad , \qquad (22)$$

where the subscript $H$ emphasizes that the Hamiltonian approach has been used in the computation.

The difference between $\mathbf{p_r}^H$ and $\mathbf{p_r}$

Indeed, the two matrices are different: $\mathbf{p_r} \equiv \mathbf{p_r}^L = L_{\mathbf{rr}}(\mathbf{x},\mathbf{r})$ is singular, with a vanishing determinant, while $\mathbf{p_r}^H = H_{\mathbf{pp}}^{-1}(\mathbf{x},\mathbf{p})$ is regular and invertible (and $H_{\mathbf{pp}}(\mathbf{x},\mathbf{p})$ as well, of course). Why? The reason is that these are two different derivatives of the slowness wrt the ray direction.

Both matrices predict the infinitesimal change of the slowness vector, given the variation of the ray velocity direction,

$$\Delta \mathbf{p} = L_{\mathbf{rr}} \Delta \mathbf{r} \quad , \quad \Delta \mathbf{p}^H = H_{\mathbf{pp}}^{-1} \Delta \mathbf{r} \qquad (23)$$

Matrix $\mathbf{p_r} = L_{\mathbf{rr}}$ has been obtained by analyzing equation set 18 of Part 1 (after this set has been solved),

$$H_\mathbf{p}(\mathbf{x},\mathbf{p}) \times \mathbf{r} = 0 \quad , \quad H(\mathbf{x},\mathbf{p}) = 0 \qquad . \qquad (24)$$

Matrix $\mathbf{p_r} = L_{\mathbf{rr}}$ takes into account that for the updated ray direction, $\mathbf{r} + \Delta \mathbf{r}$, and the updated slowness, $\mathbf{p} + \Delta \mathbf{p}$, both equations of set 24 are still satisfied,

$$H_\mathbf{p}(\mathbf{x},\mathbf{p}+\Delta\mathbf{p}) \times (\mathbf{r}+\Delta\mathbf{r}) = 0 \quad , \quad H(\mathbf{x},\mathbf{p}+\Delta\mathbf{p}) = 0 \qquad . \qquad (25)$$

The second equation of set 25, can be linearized for infinitesimal slowness variations,

$$H(\mathbf{x},\mathbf{p}+\Delta\mathbf{p}) = H(\mathbf{x},\mathbf{p}) + H_\mathbf{p} \cdot \Delta\mathbf{p} = H(\mathbf{x},\mathbf{p}) + \mathbf{r} \cdot \Delta\mathbf{p} \qquad , \qquad (26)$$

resulting in,

$$\mathbf{r} \cdot \Delta\mathbf{p} = 0 \qquad . \qquad (27)$$

This leads to the following conclusion: To preserve the constant value of the Hamiltonian, only the changes of the slowness, in the plane normal to the ray (ray-normal changes of the slowness vector), are allowed. The changes of the slowness, in the direction tangent to the ray (ray-tangent changes of the slowness vector), are prohibited by equation 27.

Now consider the inverse matrix $H_{\mathbf{pp}}^{-1}$. First, we note that,

$$H_{\mathbf{p}}(\mathbf{x},\mathbf{p}) = \mathbf{r} \quad \rightarrow \quad H_{\mathbf{p}}(\mathbf{x},\mathbf{p}) \times \mathbf{r} = 0 \quad . \tag{28}$$

Thus, the first equation of set 24 also agrees with the Hamiltonian approach. However, the Hamiltonian's Hessian matrix wrt the slowness vector, $H_{\mathbf{pp}}(\mathbf{x},\mathbf{p})$, does not take into account that the updated Hamiltonian has to remain constant,

$$H(\mathbf{x},\mathbf{p}) = 0 \quad , \quad H(\mathbf{x},\mathbf{p}+\Delta\mathbf{p}^H) \neq 0 \quad . \tag{29}$$

Vector $\Delta\mathbf{p}$ has only the ray-normal counterpart, while vector $\Delta\mathbf{p}^H$ has both ray-normal and ray-tangent counterparts; $\Delta\mathbf{p}$ is a normal projection of $\Delta\mathbf{p}^H$,

$$\Delta\mathbf{p} = \mathbf{r} \times \Delta\mathbf{p}^H \times \mathbf{r} \quad , \quad \text{or} \quad \Delta\mathbf{p} = \mathbf{T}\Delta\mathbf{p}^H \quad , \tag{30}$$

where $\mathbf{T} = \mathbf{I} - \mathbf{r} \otimes \mathbf{r}$ is the same transformation matrix (tensor) used in Part I, for example, to relate the non-normalized directional gradient of the ray velocity with the normalized one. Combining equations 23 and 30, we obtain,

$$\mathbf{T}\Delta\mathbf{p}^H = L_{\mathbf{rr}} \Delta\mathbf{r} \quad , \quad \Delta\mathbf{p}^H = H_{\mathbf{pp}}^{-1} \Delta\mathbf{r} \quad , \tag{31}$$

which leads to,

$$\mathbf{T} H_{\mathbf{pp}}^{-1} \Delta\mathbf{r} = L_{\mathbf{rr}} \Delta\mathbf{r} \quad \rightarrow \quad \left(\mathbf{T} H_{\mathbf{pp}}^{-1} - L_{\mathbf{rr}}\right)\Delta\mathbf{r} = 0 \quad . \tag{32}$$

Had the variation $\Delta\mathbf{r}$ be a vector of a fixed physical direction (like the slowness, ray direction, directional gradient of the ray velocity, etc.), this would mean that $\Delta\mathbf{r}$ is an eigenvector of the matrix in the brackets of equation 32, with the corresponding eigenvalue zero; however this is not the case. Since the ray direction vector is normalized to the unit length, $\mathbf{r}\cdot\mathbf{r}=1$, the arbitrary small vector $\Delta\mathbf{r}$ must be in the ray-normal plane,

$$\mathbf{r}\cdot\mathbf{r}=1 \quad \rightarrow \quad \mathbf{r}\cdot\Delta\mathbf{r}=0 \quad . \tag{33}$$

Furthermore, neither $L_{\mathbf{rr}}$, nor $H_{\mathbf{pp}}$ or its inverse $H_{\mathbf{pp}}^{-1}$, do not account for the "incomplete" constraint of equation 33 (that still allows for $\Delta\mathbf{r}$ any azimuth in the ray-normal plane); recall that $\partial\mathbf{p}/\partial\mathbf{r}=L_{\mathbf{rr}}$ is the *non-normalized* directional gradient of the slowness vector. Therefore, $\Delta\mathbf{r}$ in equation 32 is an arbitrary small vector, which means that the whole matrix in the brackets in this equation vanishes, and this results in,

$$L_{\mathbf{rr}}=\mathbf{T}H_{\mathbf{pp}}^{-1} \tag{34}$$

where the transform operator $\mathbf{T}$ (defined here after equation 30, and also in equation E3 Part I) can be interpreted as an operator that removes the ray-tangent counterpart of any vector to which it is applied. As we see from equation 34, operator $\mathbf{T}$ can be applied also to a second-order tensor; in this case it removes the ray-tangent counterpart a tensor, proportional (up to a scalar factor) to $\mathbf{r}\otimes\mathbf{r}$. Note that equation 34 makes it possible to establish matrix $L_{\mathbf{rr}}$, given $H_{\mathbf{pp}}$, but not vice versa: both matrices, $L_{\mathbf{rr}}$ and $\mathbf{T}$, are singular.

The similarity between $\mathbf{p}_\mathbf{r}^H$ and $\mathbf{p}_\mathbf{r}$

Despite the principal difference between matrices $L_{\mathbf{rr}}$ and $H_{\mathbf{pp}}^{-1}$, mentioned above, there is also a principal similarity: their three eigenvectors are identical, and one of them is the ray direction $\mathbf{r}$ (see Appendix C). Two of the three corresponding eigenvalues of $L_{\mathbf{rr}}$ and $H_{\mathbf{pp}}^{-1}$ are also identical, where the third eigenvalue, related to the eigenvector $\mathbf{r}$, is different in the two matrices (see, for example, equations C8 and C9 of the numerical computation for a triclinic medium). This eigenvalue is zero for matrix $L_{\mathbf{rr}}$ with the vanishing determinant, and nonzero value $\lambda_{\mathbf{r}}$ for the inverse matrix $H_{\mathbf{pp}}^{-1}$. All three eigenvalues, including $\lambda_{\mathbf{r}}$, have the units of slowness. For isotropic media, the two nonzero eigenvalues of matrix $L_{\mathbf{rr}}$ are the reciprocals of the medium velocity, $\lambda_1 = \lambda_2 = v^{-1}$. For a general anisotropic case, $\lambda_1 \neq \lambda_2$. Note that the nonzero eigenvalues of matrix $L_{\mathbf{rr},S}$ (computed at the source point), which appear in equation 29 of Part 5 relating the ray Jacobian to the (relative) geometric spreading, and their corresponding eigenvectors (in the ray-normal plane), are used to establish the initial conditions for the point-source and plane-wave basic solutions of the Jacobi DRT equation (equations 18 and 20 of Part 5, respectively).

Recall that any symmetric matrix $\mathbf{A}$ of dimension $n$ with (real) eigenvalues $\lambda_i$ and eigenvectors $\mathbf{v}_i$ can be naturally expanded by its eigensystem,

$$\mathbf{A} = \sum_{i=1}^{n} \lambda_i \mathbf{v}_i \otimes \mathbf{v}_i \quad . \tag{35}$$

Two items (matrices) in this summation, with the eigenvectors in the plane normal to the ray, are identical for both matrices, $L_{\mathbf{rr}}$ and $H_{\mathbf{pp}}^{-1}$. The third item is zero for matrix $L_{\mathbf{rr}}$, and $\lambda_{\mathbf{r}} \mathbf{r} \otimes \mathbf{r}$ for matrix $H_{\mathbf{pp}}^{-1}$. Note that there is no need to solve a cubic equation in order to find the eigenvalue $\lambda_{\mathbf{r}}$. With the reference Hamiltonian $H^{\bar{\tau}}$ or the arclength-related Hamiltonian $H$, defined in equations 11 and 13 of Part I, respectively, the eigenvalue $\lambda_{\mathbf{r}}$ becomes,

$$\lambda_{\mathbf{r}}^{-1} = \mathbf{r} \cdot H_{\mathbf{pp}} \cdot \mathbf{r} \quad , \quad \lambda_{\mathbf{r}}^{-1} = -\frac{H_{\mathbf{p}}^{\bar{\tau}} \cdot H_{\mathbf{pp}}^{\bar{\tau}} \cdot H_{\mathbf{p}}^{\bar{\tau}}}{\left(H_{\mathbf{p}}^{\bar{\tau}} \cdot H_{\mathbf{p}}^{\bar{\tau}}\right)^{3/2}} \quad . \tag{36}$$

As mentioned, operator $\mathbf{T}$ wipes out the ray-tangent component (unnecessary with the Lagrangian approach), and therefore, the eigenvalue $\lambda_{\mathbf{r}}$ of the Hamiltonian's inverse Hessian $H_{\mathbf{pp}}^{-1}$ (related to the eigenvector $\mathbf{r}$) cannot be restored from the Lagrangian derivatives alone. The eigenvectors of the Hessian, $H_{\mathbf{pp}}$, and its inverse, $H_{\mathbf{pp}}^{-1}$, are identical (while their corresponding eigenvalues are reciprocals to each other). Since matrices $L_{\mathbf{rr}}$ and $H_{\mathbf{pp}}$ have the same eigenvectors, they can be simultaneously diagonalized. Consequently, their product is commutative (e.g., Anton, 1987; Horn and Johnson, 2013),

$$L_{\mathbf{rr}} H_{\mathbf{pp}} = H_{\mathbf{pp}} L_{\mathbf{rr}} = \mathbf{T} \quad . \tag{37}$$

### SLOWNESS OF PLANE-WAVE PARAXIAL RAYS AT THE SOURCE

Our definition of plane-wave paraxial rays differs from the standard one. We call plane-wave paraxial rays those, whose ray direction at the start point is collinear to that of the central ray. This definition is suitable for the Lagrangian formulation, where the primary DoF are the ray location and directions. The standard formulations assume collinear slowness directions of the paraxial plane-wave and central rays as it is mainly oriented on the Hamiltonian approach, where the primary DoF are the ray location and slowness vectors. In Appendix D, we study the difference between the two slowness directions that follows from our modified definition.

## CONCLUSIONS

Considering a general wave type, we demonstrate that both Cartesian DRT formulations: the proposed Jacobi Lagrangian-based second-order ODE, and the commonly used Hamiltonian-based first-order ODE set, are fully consistent for the computation of the dynamic ray properties in isotropic and general anisotropic media.

We provide the fundamental connection between the Lagrangian's directional Hessian and the inverse matrix of the Hamiltonian's slowness Hessian; we compare their eigensystems and we interpret the physical nature of the differences between the two matrices. We demonstrate the two approaches numerically, using an example of spatially varying triclinic medium, and analytically, for an ellipsoidal orthorhombic medium.

We show that both, the proposed Lagrangian-based and the Hamiltonian-based DRT solutions have identical ray-normal vector components. In addition, the Hamiltonian solution is accompanied by an inessential and unnecessary ray-tangent counterpart that does not affect the ray Jacobian and the geometric spreading. We therefore consider the Lagrangian approach simpler, but both approaches are elsewise similar, and the final choice of the method may depend on the specific DRT problem.

## ACKNOWLEDGEMENT

The authors are grateful to Emerson for financial and technical support for this study and for permission to publish its results. The gratitude is extended to Ivan Pšenčík, Einar Iversen, Michael Slawinski, Alexey Stovas, Vladimir Grechka, and our colleague Beth Orshalimy, whose valuable remarks helped to improve the content and style of this paper.

# APPENDIX A. LAGRANGIAN VERSUS HAMILTONIAN DYNAMIC RAY TRACING APPROACH FOR ISOTROPIC MEDIA

In this appendix, we demonstrate that the proposed Lagrangian-based Jacobi DRT equation is consistent with the conventional Hamiltonian approach. We start by studying isotropic media (in this appendix), and we then consider general anisotropy (in Appendix H). To derive the conventional Hamiltonian-based DRT equations vs. the arclength, we follow Červený (2000). For the purpose of derivation, the most suitable flow parameter is $\sigma$, so we obtain the isotropic DRT equation by two ways:

- Deriving the Hamiltonian DRT equation set consisting of two first-order equations wrt $\sigma$ and then converting it into the second-order Jacobi DRT equation wrt the arclength $s$.

- Working directly with the arclength $s$.

General workflow for deriving the Hamiltonian DRT equations

Assume vector $\mathbf{p}_{\text{prx}}(\gamma, s)$ is the slowness of a paraxial ray, $v(\mathbf{x}_{\text{prx}})$ is its velocity, $\mathbf{x}_{\text{prx}}(\gamma, s)$ is the paraxial ray path, and $\gamma$ is an arbitrary (one of the four) ray coordinate (RC) defined at the source. The paraxial ray path may be presented, for example, as

$$\mathbf{x}_{\text{prx}}(\gamma, s) = \mathbf{x}(s) + \gamma \mathbf{w}(s) \quad , \tag{A1}$$

where $\mathbf{x}(s)$ is the central ray position, but this is not a must. For simplicity, we omit the index of $\gamma$ here. The three other RC are assumed fixed.

The isotropic Hamiltonian can be written as (Červený, 2000) (equation 3.1.15),

$$H_{\text{iso}}^{(n)}(\mathbf{x}_{\text{prx}}, \mathbf{p}_{\text{prx}}) = \frac{v^{2-n}(\mathbf{x})}{2} \left[ \mathbf{p}_{\text{prx}} \cdot \mathbf{p}_{\text{prx}} - v^{-2}(\mathbf{x}_{\text{prx}}) \right] \quad , \tag{A2}$$

where $n$ is the index of the flow parameter: for the traveltime $n = 0$, for the arclength, $n = 1$, and for sigma, $n = 2$.

The DRT set is given (in our notations) by (Červený, 2000) (equation 4.2.4),

$$\frac{d}{d\varsigma^{(n)}} \begin{bmatrix} \mathbf{x}_{\text{prx},\gamma} \\ \mathbf{p}_{\text{prx},\gamma} \end{bmatrix} = \begin{bmatrix} \mathbf{A}^{(n)} & \mathbf{B}^{(n)} \\ -\mathbf{C}^{(n)} & -\mathbf{D}^{(n)} \end{bmatrix} \begin{bmatrix} \mathbf{x}_{\text{prx},\gamma} \\ \mathbf{p}_{\text{prx},\gamma} \end{bmatrix} \quad , \tag{A3}$$

where $\varsigma^{(n)}$ is a general flow parameter, in particular, for

$(\text{traveltime}: n = 0 \rightarrow \varsigma = \tau); \; (\text{arclength}: n = 1 \rightarrow \varsigma = s); \; (\text{sigma}: n = 2 \rightarrow \varsigma = \sigma)$, and

$$\mathbf{w} \equiv \mathbf{x}_{\text{prx},\gamma} = \frac{\partial \mathbf{x}_{\text{prx}}(\gamma, \varsigma)}{\partial \gamma} \quad , \quad \mathbf{p}_{\text{prx},\gamma} = \frac{\partial \mathbf{p}_{\text{prx}}(\gamma, \varsigma)}{\partial \gamma} \quad . \tag{A4}$$

The matrix in equation A3 is of dimension $6 \times 6$, consisting of four $3 \times 3$ blocks (Červený, 2000) (equation 4.2.5),

$$\mathbf{A}^{(n)} = \frac{\partial^2 H_{\text{iso}}^{(n)}(\mathbf{x}_{\text{prx}}, \mathbf{p}_{\text{prx}})}{\partial \mathbf{p}_{\text{prx}} \partial \mathbf{x}_{\text{prx}}} = H_{\text{iso, px}}^{(n)} \quad , \quad \mathbf{B}^{(n)} = \frac{\partial^2 H_{\text{iso}}^{(n)}(\mathbf{x}, \mathbf{p})}{\partial \mathbf{p}_{\text{prx}}^2} = H_{\text{iso, pp}}^{(n)} \quad ,$$

$$\mathbf{C}^{(n)} = \frac{\partial^2 H_{\text{iso}}^{(n)}(\mathbf{x}_{\text{prx}}, \mathbf{p}_{\text{prx}})}{\partial \mathbf{x}_{\text{prx}}^2} = H_{\text{iso, xx}}^{(n)} \quad , \quad \mathbf{D}^{(n)} = \frac{\partial^2 H_{\text{iso}}^{(n)}(\mathbf{x}, \mathbf{p})}{\partial \mathbf{x}_{\text{prx}} \partial \mathbf{p}_{\text{prx}}} = H_{\text{iso, xp}}^{(n)} \quad .$$

(A5)

The Hamiltonian vanishes for any value of the RC $\gamma$. We apply the chain rule, and this leads to the following constraint (Červený, 2000) (equation 4.2.7),

$$\frac{\partial H^{\varsigma}}{\partial \gamma} = \frac{\partial H^{\varsigma}}{\partial \mathbf{x}_{\text{prx}}} \cdot \frac{\partial \mathbf{x}_{\text{prx}}}{\partial \gamma} + \frac{\partial H^{\varsigma}}{\partial \mathbf{p}_{\text{prx}}} \cdot \frac{\partial \mathbf{p}_{\text{prx}}}{\partial \gamma} = \frac{\partial H^{\varsigma}}{\partial \mathbf{x}_{\text{prx}}} \cdot \mathbf{w} + \frac{\partial H^{\varsigma}}{\partial \mathbf{p}_{\text{prx}}} \cdot \mathbf{p}_{\text{prx}, \gamma} = 0 \quad , \tag{A6}$$

where $H^{\varsigma}$ is a general Hamiltonian. Hence, due to the constraint A6, only five of the six equations in set A3 are independent.

In this appendix, we apply the Hamiltonians for isotropic media, $H^{\varsigma} \to H_{\text{iso}}^{(n)}$. Note, however, that equations A3, A5 and A6 are valid for general anisotropic media as well.

DRT with respect to sigma converted to DRT with respet to the arclength

The sigma-related Hamiltonian reads,

$$H_{\text{iso}}^{(2)}(\mathbf{x}_{\text{prx}}, \mathbf{p}_{\text{prx}}) = \frac{1}{2} \left[ \mathbf{p}_{\text{prx}} \cdot \mathbf{p}_{\text{prx}} - v^{-2}(\mathbf{x}_{\text{prx}}) \right] \quad . \tag{A7}$$

The gradients of the Hamiltonian are,

$$\frac{\partial H_{\text{iso}}^{(2)}}{\partial \mathbf{x}_{\text{prx}}} = v^{-3} \nabla v \quad , \quad \frac{\partial H^{(2)}}{\partial \mathbf{p}_{\text{prx}}} = \frac{\mathbf{r}}{v} \quad , \tag{A8}$$

where $\mathbf{r}$ is the normalized ray direction (recall that for isotropic media, $v\mathbf{p} = \mathbf{r}$). The above derivatives are computed at the central ray, $\gamma = 0$.

This leads to the following constraint,

$$\nabla v \cdot \mathbf{w} + v^2 \mathbf{r} \cdot \mathbf{p}_{\text{prx},\gamma} = 0 \quad . \tag{A9}$$

The sub-matrices read,

$$H^{(2)}_{\text{iso, px}} = H^{(2)}_{\text{iso, xp}} = 0 \,, \quad H^{(2)}_{\text{iso, pp}} = \mathbf{I} \,, \quad H^{(2)}_{\text{iso, xx}} = -\frac{1}{2} \nabla \nabla v^{-2} \,, \tag{A10}$$

where $\mathbf{I}$ is the identity matrix, and the Hessian matrix $\nabla \nabla v^{-2}$ can be explicitly written as,

$$\nabla \nabla v^{-2} = 6 v^{-4} \nabla v \otimes \nabla v - 2 v^{-3} \nabla \nabla v \quad . \tag{A11}$$

The governing set A3 becomes,

$$\frac{d\mathbf{w}}{d\sigma} = \mathbf{p}_{\text{prx},\gamma} \qquad \frac{d\mathbf{p}_{\text{prx},\gamma}}{d\sigma} = \left( 3 v^{-4} \nabla v \otimes \nabla v - v^{-3} \nabla \nabla v \right) \mathbf{w} \quad . \tag{A12}$$

Note that $d\sigma = v(s) ds$, and thus, the DRT set may be arranged as,

$$\frac{\dot{\mathbf{w}}}{v} = \mathbf{p}_{\text{prx},\gamma} \,, \quad \dot{\mathbf{p}}_{\text{prx},\gamma} = \frac{3 \nabla v \otimes \nabla v - v \nabla \nabla v}{v^3} \mathbf{w} \quad . \tag{A13}$$

Finally, we eliminate $\mathbf{p}_{\text{prx},\gamma}$ from the constraint A9 and the dynamic equation set A13, and we obtain the vector-form, second-order Hamiltonian DRT equation with the constraint,

$$\frac{d}{ds}\frac{\dot{\mathbf{w}}}{v} = \frac{3\nabla v \otimes \nabla v - v\nabla\nabla v}{v^3}\mathbf{w} \quad , \qquad \frac{\nabla v}{v}\cdot\mathbf{w} + \mathbf{r}\cdot\dot{\mathbf{w}} = 0 \qquad . \tag{A14}$$

DRT with respect to the arclength

In this case, the Hamiltonian reads,

$$H^{(1)}_{\text{iso}}\left(\mathbf{x}_{\text{prx}}, \mathbf{p}_{\text{prx}}\right) = \frac{v(\mathbf{x}_{\text{prx}})}{2}\left[\mathbf{p}_{\text{prx}}\cdot\mathbf{p}_{\text{prx}} - v^{-2}(\mathbf{x}_{\text{prx}})\right] \qquad . \tag{A15}$$

The gradients of the Hamiltonian are,

$$\frac{\partial H^{(1)}_{\text{iso}}}{\partial \mathbf{x}_{\text{prx}}} = v^{-2}\nabla v \quad , \qquad \frac{\partial H^{(1)}_{\text{iso}}}{\partial \mathbf{p}_{\text{prx}}} = \mathbf{r} \qquad . \tag{A16}$$

The constraint coincides with equation A9. The sub-matrices read,

$$H^{(1)}_{\text{iso, px}} = \frac{\mathbf{r}\otimes\nabla v}{v} \quad , \qquad H^{(1)}_{\text{iso, pp}} = v\mathbf{I} \quad ,$$
$$H^{(1)}_{\text{iso, xx}} = \frac{v\nabla\nabla v - \nabla v \otimes \nabla v}{v^3} \quad , \quad H^{(1)}_{\text{iso, xp}} = \frac{\nabla v \otimes \mathbf{r}}{v} \quad . \tag{A17}$$

The governing set A3 becomes,

$$\dot{\mathbf{w}} = \frac{\mathbf{r}\otimes\nabla v}{v}\cdot\mathbf{w} + v\mathbf{p}_{\text{prx},\gamma} \quad , \qquad \dot{\mathbf{p}}_{\text{prx},\gamma} = \frac{\nabla v \otimes \nabla v - v\nabla\nabla v}{v^3}\cdot\mathbf{w} - \frac{\nabla v \otimes \mathbf{r}}{v}\cdot\mathbf{p}_{\text{prx},\gamma} \qquad . \tag{A18}$$

Eliminating vector $\mathbf{p}_{\text{prx},\gamma}$ from the first equation of set A18,

$$\frac{\dot{\mathbf{w}}}{v} - \frac{\mathbf{r}\otimes\nabla v}{v^2}\cdot\mathbf{w} = \mathbf{p}_{\text{prx},\gamma} \qquad . \tag{A19}$$

Differentiating equation A19 and introducing the second equation of set A18, we obtain,

$$\frac{d}{ds}\left(\frac{\dot{\mathbf{w}}}{v} - \frac{\mathbf{r}\otimes\nabla v}{v^2}\cdot\mathbf{w}\right) = \frac{\nabla v\otimes\nabla v - v\nabla\nabla v}{v^3}\cdot\mathbf{w} - \frac{\nabla v\otimes\mathbf{r}}{v}\cdot\mathbf{p}_{\text{prx},\gamma} \qquad . \tag{A20}$$

We introduce equation A19 into the last term of A20,

$$\frac{d}{ds}\left(\frac{\dot{\mathbf{w}}}{v} - \frac{\mathbf{r}\otimes\nabla v}{v^2}\cdot\mathbf{w}\right) = \frac{\nabla v\otimes\nabla v - v\nabla\nabla v}{v^3}\cdot\mathbf{w} - \frac{\nabla v\otimes\mathbf{r}}{v}\cdot\frac{\dot{\mathbf{w}}}{v} + \frac{(\nabla v\otimes\mathbf{r})\cdot(\mathbf{r}\otimes\nabla v)}{v^3}\cdot\mathbf{w} \qquad . \tag{A21}$$

Note that for any four vectors $\mathbf{a}$, $\mathbf{b}$, $\mathbf{c}$, $\mathbf{d}$, the following identity holds,

$$(\mathbf{a}\otimes\mathbf{b})\cdot(\mathbf{c}\otimes\mathbf{d}) = (\mathbf{b}\cdot\mathbf{c})\mathbf{a}\otimes\mathbf{d} \qquad . \tag{A22}$$

A particular case of this identity reads,

$$(\mathbf{a}\otimes\mathbf{b})\cdot(\mathbf{b}\otimes\mathbf{a}) = (\mathbf{b}\cdot\mathbf{b})\mathbf{a}\otimes\mathbf{a} \qquad . \tag{A23}$$

Taking into account that $\mathbf{r}\cdot\mathbf{r}=1$, we apply this property to the last term of equation A21, and the equation simplifies to the first equation of set A24,

$$\frac{d}{ds}\left(\frac{\dot{\mathbf{w}}}{v} - \frac{\mathbf{r}\otimes\nabla v}{v^2}\cdot\mathbf{w}\right) = \frac{2\nabla v\otimes\nabla v - v\nabla\nabla v}{v^3}\cdot\mathbf{w} - \frac{\nabla v\otimes\mathbf{r}}{v}\cdot\frac{\dot{\mathbf{w}}}{v} \quad , \quad \dot{\mathbf{w}}\cdot\mathbf{r}=0 \qquad . \tag{A24}$$

Combining equations A9 and A19, we eliminate $\mathbf{p}_{\text{prx},\gamma}$ from the constraint,

$$\nabla v\cdot\mathbf{w} + v^2\mathbf{r}\cdot\left(\frac{\dot{\mathbf{w}}}{v} - \frac{\mathbf{r}\otimes\nabla v}{v^2}\cdot\mathbf{w}\right) = 0 \qquad . \tag{A25}$$

Application of the auxiliary equation,

$$(\mathbf{a}\otimes\mathbf{b})\mathbf{c} = (\mathbf{a}\otimes\mathbf{c})\mathbf{b} = (\mathbf{b}\cdot\mathbf{c})\mathbf{a} \quad , \tag{A26}$$

where $\mathbf{a}, \mathbf{b}, \mathbf{c}$ are arbitrary vectors, to the second term in the brackets of equation A25, leads to the second equation (constraint) of set A23, $\dot{\mathbf{w}} \cdot \mathbf{r} = 0$.

Comparison with the proposed Jacobi DRT equation

The Jacobi DRT equation 12 (accompanied by the "normal" constraint of equation 13) includes the coefficient matrices $L_{\mathbf{xx}}$, $L_{\mathbf{xr}}$, $L_{\mathbf{rx}}$ and $L_{\mathbf{rr}}$ listed in equation set A2. For an isotropic case, these matrices simplify to,

$$L_{\mathbf{xx}} = \frac{2\nabla v \otimes \nabla v}{v^3} - \frac{\nabla \nabla v}{v^2} \quad , \quad L_{\mathbf{xr}} = -\frac{\nabla v \otimes \mathbf{r}}{v^2} \quad , \quad L_{\mathbf{rx}} = -\frac{\mathbf{r} \otimes \nabla v}{v^2} \quad , \quad L_{\mathbf{rr}} = \frac{\mathbf{I} - \mathbf{r} \otimes \mathbf{r}}{v} \quad . \tag{A27}$$

Introduction of equation A27 into 12 leads to,

$$\frac{d}{ds}\left(\frac{\mathbf{I}-\mathbf{r}\otimes\mathbf{r}}{v} \cdot \dot{\mathbf{u}} - \frac{\mathbf{r}\otimes\nabla v}{v^2} \cdot \mathbf{u}\right) = \frac{2\nabla v \otimes \nabla v - v \nabla \nabla v}{v^3} \cdot \mathbf{u} - \frac{\nabla v \otimes \mathbf{r}}{v^2} \cdot \dot{\mathbf{u}} \quad , \quad \mathbf{u} \cdot \mathbf{r} = 0 \quad , \tag{A28}$$

where $\mathbf{u}$ is the solution of the Jacobi equation, $\mathbf{u}(s) \neq \mathbf{w}(s)$. Recall that the constraint (the second equation of set A28, "the normal solution") is not an inherent property of the Jacobi DRT equation. The Jacobi DRT equation only says that the tangent part of the solution is undefined (and may be arbitrary). Since the component tangent to the ray has no effect on the ray Jacobian and the geometric spreading, we set it to zero in the solution of the Jacobi DRT.

Next, we assume that the Lagrangian and the Hamiltonian DRT solutions, $\mathbf{u}(s)$ and $\mathbf{w}(s)$ differ by the tangent component only,

$$\mathbf{w}(s) = \mathbf{u}(s) + \rho_t(s)\mathbf{r}(s) \quad , \tag{A29}$$

where $\rho_t(s)$ is a scalar function, and we demonstrate that with this assumption the two approaches lead to identical, linear, second-order ODEs. Indeed, as mentioned above, the three Cartesian components of the Jacobi equation are dependent, and therefore, the Jacobi equation is insensitive to the tangent counterpart of the solution. In other words, if the normal shift $\mathbf{u}(s)$ is a solution of the Jacobi equation, then vector $\mathbf{w}(s)$ defined in equation A27 is its solution as well for any scalar function $\rho_t(s)$. Thus, $\mathbf{u}(s)$ can be replaced by $\mathbf{w}(s)$ in the first equation of set A28,

$$\frac{d}{ds}\left(\frac{\mathbf{I}-\mathbf{r}\otimes\mathbf{r}}{v}\cdot\dot{\mathbf{w}}-\frac{\mathbf{r}\otimes\nabla v}{v^2}\cdot\mathbf{w}\right)=\frac{2\nabla v\otimes\nabla v-v\nabla\nabla v}{v^3}\cdot\mathbf{w}-\frac{\nabla v\otimes\mathbf{r}}{v^2}\cdot\dot{\mathbf{w}} \quad , \qquad (A30)$$

We now compare the Lagrangian and Hamiltonian DRT equations (where the latter was obtained wrt sigma and then converted for the flow parameter arclength). The "residual" (difference) between equations A30 and A14 reads,

$$-\frac{d}{ds}\left(\frac{\mathbf{r}\otimes\mathbf{r}}{v}\cdot\dot{\mathbf{w}}+\frac{\mathbf{r}\otimes\nabla v}{v^2}\cdot\mathbf{w}\right)=-\frac{\nabla v\otimes\nabla v}{v^3}\cdot\mathbf{w}-\frac{\nabla v\otimes\mathbf{r}}{v^2}\cdot\dot{\mathbf{w}} \quad , \qquad \frac{\nabla v}{v}\cdot\mathbf{w}+\mathbf{r}\cdot\dot{\mathbf{w}}=0 \quad , \qquad (A31)$$

where the second equation of set A31 is the constraint. Applying equation A26, the residual is simplified to,

$$-\frac{d}{ds}\left(\frac{\nabla v\cdot\mathbf{w}+v\mathbf{r}\cdot\dot{\mathbf{w}}}{v^2}\mathbf{r}\right)=-\frac{\nabla v\cdot\mathbf{w}+v\mathbf{r}\cdot\dot{\mathbf{w}}}{v^3}\nabla v \quad . \qquad (A32)$$

Both, the left- and right-hand sides of equation A32 vanish separately due to the constraint.

Next, we compare the Lagrangian and Hamiltonian DRT equations, where the latter was obtained immediately wrt the arclength. The residual between equations A30 and A24 reads,

$$-\frac{d}{ds}\left(\frac{\mathbf{r}\otimes\mathbf{r}}{v}\cdot\dot{\mathbf{w}}\right)=0 \quad, \quad \dot{\mathbf{w}}\cdot\mathbf{r}=0 \quad . \tag{A33}$$

Applying equation A26, the residual is simplified to,

$$-\frac{d}{ds}\left(\frac{\mathbf{r}\cdot\dot{\mathbf{u}}}{v}\mathbf{r}\right)=0 \quad, \tag{A34}$$

where the numerator vanishes due to the constraint (the second equation of set A33).

In summary, we conclude that equation A30 can be applied for all three cases: the proposed Jacobi DRT equation and the two Hamiltonian approaches (with the arclength $s$ used directly as the flow parameter, and with $\sigma$ converted then to $s$). However, the constraints are different,

$$\underbrace{\begin{array}{c}\rho_{t,S}=0 \quad \rightarrow \quad \mathbf{w}=\mathbf{u}, \\ \mathbf{u}\cdot\mathbf{r}=0, \; \dot{\mathbf{u}}\cdot\mathbf{r}+\mathbf{u}\cdot\dot{\mathbf{r}}=0,\end{array}}_{\text{Jacobi constraint}} \quad \underbrace{\dot{\mathbf{w}}\cdot\mathbf{r}=0}_{\text{Červený constraint, }s} \quad, \quad \underbrace{\frac{\nabla v}{v}\cdot\mathbf{w}+\dot{\mathbf{w}}\cdot\mathbf{r}=0}_{\text{Červený constraint, }\sigma\rightarrow s} \quad . \tag{A35}$$

Equation A14 with its constraint is equivalent to A30. Equation A24 with its constraint is also equivalent to A30. Equations A14 and A24, including their corresponding constraints, are equivalent to each other. These different constraints lead to the same geometric spreading.

Note also that both the Jacobi normal solution $\mathbf{u}$ and the Hamiltonian solution $\mathbf{w}$ vanish at the origin of the point-source ray; thus, the three constraints in equation A35 become identical and reduce to, $\dot{\mathbf{u}}_S\cdot\mathbf{r}_S=0$ or $\dot{\mathbf{w}}_S\cdot\mathbf{r}_S=0$, where subscript $S$ means the source point.

# APPENDIX B. LAGRANGIAN VERSUS HAMILTONIAN DYNAMIC RAY TRACING APPROACH FOR GENERAL ANISOTROPIC MEDIA

In this appendix, we compare the DRT equations that follow from the Hamiltonian and Lagrangian approaches for general anisotropic media, with the flow parameter arclength. Recall that we use the rules of tensor algebra and do not distinguish between row and column vectors.

The Hamiltonian DRT equation set can be arranged as (e.g., Červený, 2000) (equation 4.2.4),

$$\dot{\mathbf{w}} = H_{\mathbf{px}} \mathbf{w} + H_{\mathbf{pp}} \mathbf{p}_{\text{prx},\gamma} \quad , \quad \dot{\mathbf{p}}_{\text{prx},\gamma} = -H_{\mathbf{xx}} \mathbf{w} - H_{\mathbf{xp}} \mathbf{p}_{\text{prx},\gamma} \quad , \tag{B1}$$

where, in our case, $H \equiv H^s$ is the arclength-related Hamiltonian, defined in equations 11 and 13 of Part I. We emphasize that the relationships in this appendix are only valid for this Hamiltonian.

We first eliminate $\mathbf{p}_{\text{prx},\gamma}$. From the first equation of set B1 we obtain,

$$\mathbf{p}_{\text{prx},\gamma} = H_{\mathbf{pp}}^{-1} \dot{\mathbf{w}} - H_{\mathbf{pp}}^{-1} H_{\mathbf{px}} \mathbf{w} \quad , \quad \dot{\mathbf{p}}_{\text{prx},\gamma} = \frac{d}{ds}\left( H_{\mathbf{pp}}^{-1} \dot{\mathbf{w}} - H_{\mathbf{pp}}^{-1} H_{\mathbf{px}} \mathbf{w} \right) \quad . \tag{B2}$$

We combine this result with the second equation of set B1,

$$\frac{d}{ds}\left( H_{\mathbf{pp}}^{-1} \dot{\mathbf{w}} - H_{\mathbf{pp}}^{-1} H_{\mathbf{px}} \mathbf{w} \right) = -H_{\mathbf{xx}} \mathbf{w} - H_{\mathbf{xp}} \mathbf{p}_{\text{prx},\gamma} \quad . \tag{B3}$$

Next, we use the first equation of set B2, and this leads to the Hamiltonian DRT in terms of the solution $\mathbf{w}$ and its derivatives,

$$\frac{d}{ds}\left( H_{\mathbf{pp}}^{-1} \dot{\mathbf{w}} - H_{\mathbf{pp}}^{-1} H_{\mathbf{px}} \mathbf{w} \right) = -H_{\mathbf{xp}} H_{\mathbf{pp}}^{-1} \dot{\mathbf{w}} + \left( H_{\mathbf{xp}} H_{\mathbf{pp}}^{-1} H_{\mathbf{px}} - H_{\mathbf{xx}} \right) \mathbf{w} \quad . \tag{B4}$$

This is the second-order Hamiltonian DRT equation. It has been obtained by combining the two first-order Hamiltonian DRT equations, with the subsequent elimination of vector $\mathbf{p}_{\text{prx},\gamma}$ and its derivative wrt the arclength of the central ray. In addition to this vector equation, there is a scalar constraint (Červený, 2000) (equation 4.2.7),

$$\frac{dH}{d\gamma} = H_{\mathbf{x}}\mathbf{w} + H_{\mathbf{p}}\mathbf{p}_{\text{prx},\gamma} = 0 \qquad . \tag{B5}$$

With the first equation of set B2, we eliminate $\mathbf{p}_{\text{prx},\gamma}$, and the constraint becomes,

$$H_{\mathbf{p}} H_{\mathbf{pp}}^{-1} \cdot \dot{\mathbf{w}} + \left( H_{\mathbf{x}} - H_{\mathbf{p}} H_{\mathbf{pp}}^{-1} H_{\mathbf{px}} \right) \cdot \mathbf{w} = 0 \qquad . \tag{B6}$$

For a general heterogeneous anisotropic medium, with all stiffness components varying in 3D space, the vector in brackets vanishes,

$$H_{\mathbf{x}} = H_{\mathbf{p}} H_{\mathbf{pp}}^{-1} H_{\mathbf{px}} \quad , \quad \text{or} \quad H_{\mathbf{x}} = H_{\mathbf{xp}} H_{\mathbf{pp}}^{-1} H_{\mathbf{p}} \quad , \tag{B7}$$

and the Hamiltonian DRT constraint simplifies to,

$$\dot{\mathbf{w}} H_{\mathbf{pp}}^{-1} H_{\mathbf{p}} = 0 \quad , \quad \text{or} \quad \dot{\mathbf{w}} H_{\mathbf{pp}}^{-1} \mathbf{r} = 0 \qquad . \tag{B8}$$

Lemma

We prove the identity of equation B7. First, we note that,

$$H_{\mathbf{x}} = -\dot{\mathbf{p}} \quad , \qquad H_{\mathbf{p}} = \dot{\mathbf{x}} = \mathbf{r} \quad ,$$

$$H_{\mathbf{xp}} = \frac{\partial H_{\mathbf{x}}}{\partial \mathbf{p}} = -\frac{\partial \dot{\mathbf{p}}}{\partial \mathbf{p}} \quad , \quad H_{\mathbf{pp}}^{-1} = \left( \frac{\partial H_{\mathbf{p}}}{\partial \mathbf{p}} \right)^{-1} \left( \frac{\partial \mathbf{r}}{\partial \mathbf{p}} \right)^{-1} = \left. \frac{\partial \mathbf{p}}{\partial \mathbf{r}} \right|_H \quad . \tag{B9}$$

The subscript $H$ in the last equation of set B9 emphasize that the slowness (directional) gradient tensor, $\partial \mathbf{p} / \partial \mathbf{r}|_H$, has been computed with the Hamiltonian approach, rather than with the Lagrangian. Introduction of equation set B9 into B7 shows that the relationship we need to validate reads,

$$-\dot{\mathbf{p}} = -\frac{\partial \dot{\mathbf{p}}}{\partial \mathbf{p}} \frac{\partial \mathbf{p}}{\partial \mathbf{r}}\bigg|_H \mathbf{r} \quad , \tag{B10}$$

and after the reduction of the chain rule, it simplifies to,

$$\frac{\partial \dot{\mathbf{p}}}{\partial \mathbf{r}} \mathbf{r} = \dot{\mathbf{p}} \quad . \tag{B11}$$

Equation B11 is Euler's theorem for a first-degree homogeneous function wrt the ray direction vector. If we prove that the arclength derivative $\dot{\mathbf{p}}$, is by definition, such a function, then the property of equation B11 holds, and as a result, equations B10 and B7 also hold. Thus, we need to explore the sensitivity of vector $\dot{\mathbf{p}}$ to the length $k$ of the tangent vector $k\mathbf{r}$. For this, we assume the flow parameter has changed from the actual arclength, $s$, to the scaled arclength, $s^* = s/k$. As explained in Part II, the slowness vector itself is a physical parameters insensitive to the length $k$. However, this is not so for its derivative wrt the flow parameter,

$$\dot{\mathbf{p}}_{s^*}(k\mathbf{r}) = \frac{\partial \mathbf{p}}{\partial s^*} = \frac{\partial \mathbf{p}}{\partial s/k} = k \frac{\partial \mathbf{p}}{\partial s} = k \dot{\mathbf{p}}(\mathbf{r}) \quad , \quad \text{where} \quad k\mathbf{r} = \frac{d\mathbf{x}}{ds^*} = k \frac{d\mathbf{x}}{ds} \quad . \tag{B12}$$

Equation B12 finalizes the proof of the lemma: $\dot{\mathbf{p}}(\mathbf{r})$ is a first-degree homogeneous function by the definition of such function. Identity B7 follows from this proof.

Note that the ray velocity direction **r** is one of the eigenvectors for the Hamiltonian Hessian matrix wrt the slowness components, $H_{\mathbf{pp}}$ (and thus also for its inverse, $H_{\mathbf{pp}}^{-1}$). Denoting by $\lambda_{\mathbf{r}}$ the corresponding eigenvalue for the inverse matrix $H_{\mathbf{pp}}^{-1}$, we obtain,

$$H_{\mathbf{pp}}^{-1}\mathbf{r} = \lambda_{\mathbf{r}}\mathbf{r}, \quad \lambda_{\mathbf{r}} \neq 0 \tag{B13}$$

and the constraint further simplifies to,

$$\dot{\mathbf{w}}\, H_{\mathbf{pp}}^{-1}\mathbf{r} = \dot{\mathbf{w}} \cdot \lambda_{\mathbf{r}}\,\mathbf{r} = 0\,, \quad \lambda_{\mathbf{r}} \neq 0\,, \quad \rightarrow \quad \dot{\mathbf{w}} \cdot \mathbf{r} = 0 \quad , \tag{B14}$$

i.e. the constraint is the same as for an isotropic case. The Hamiltonian property in equation B7 simplifies to,

$$H_{\mathbf{x}} = \lambda_{\mathbf{r}} H_{\mathbf{xp}} H_{\mathbf{p}} \quad . \tag{B15}$$

Recall that the (Lagrangian-based) Jacobi DRT equation reads,

$$\frac{d}{ds}\left(L_{\mathbf{rr}} \cdot \dot{\mathbf{u}} + L_{\mathbf{rx}} \cdot \mathbf{u}\right) = L_{\mathbf{xr}} \cdot \dot{\mathbf{u}} + L_{\mathbf{xx}} \cdot \mathbf{u} \quad . \tag{B16}$$

As mentioned, the Jacobi DRT equation is insensitive to the tangent counterpart of the solution, and the normal shift $\mathbf{u}(s)$ can be replaced by the Hamiltonian solution $\mathbf{w}(s)$ defined in equation G29,

$$\frac{d}{ds}\left(L_{\mathbf{rr}} \cdot \dot{\mathbf{w}} + L_{\mathbf{rx}} \cdot \mathbf{w}\right) = L_{\mathbf{xr}} \cdot \dot{\mathbf{w}} + L_{\mathbf{xx}} \cdot \mathbf{w} \quad . \tag{B17}$$

Equation B17 looks very similar to the Hamiltonian equation B4, but there is no full equivalence of all coefficients, until the constraint B14 is taken into account. Note also that matrix $H_{pp}$ is invertible, while $L_{rr}$ is not, $\det L_{rr} = 0$.

We multiply vector $H_p = r$ by the vanishing scalar product $\lambda_r \dot{w} \cdot r$, and with the use of the auxiliary equation G26, we re-arrange it as,

$$\lambda_r (\dot{w} \cdot r) r = \lambda_r (r \otimes r) \dot{w} \quad . \tag{B18}$$

Next, we subtract the vanishing expression in equation B18 from the expression in brackets on the left side of the Hamiltonian DRT set B4. This leads to,

$$\frac{d}{ds}\left[\left(H_{pp}^{-1} - \lambda_r r \otimes r\right)\dot{w} - H_{pp}^{-1} H_{px} w\right] = -H_{xp} H_{pp}^{-1} \dot{w} + \left(H_{xp} H_{pp}^{-1} H_{px} - H_{xx}\right) w \quad . \tag{B19}$$

Now there is a full match of all items in the Hamiltonian and the (Lagrangian) Jacobi DRT equations. The second derivatives of the Lagrangian and the Hamiltonian are related by,

$$\begin{aligned}
L_{rr} &= H_{pp}^{-1} - \lambda_r r \otimes r \quad, & L_{rx} &= -H_{pp}^{-1} H_{px} \quad, \\
L_{xr} &= -H_{xp} H_{pp}^{-1} \quad, & L_{xx} &= H_{xp} H_{pp}^{-1} H_{px} - H_{xx} \quad,
\end{aligned} \tag{B20}$$

and the inverse relationships are obvious, provided the eigenvalue $\lambda_r$ has been computed,

$$\begin{aligned}
H_{pp}^{-1} &= L_{rr} + \lambda_r r \otimes r \quad, & H_{px} &= -H_{pp} L_{rx} \quad, \\
H_{xp} &= -L_{xr} H_{pp} \quad, & H_{xx} &= H_{xp} H_{pp}^{-1} H_{px} - L_{xx} \quad.
\end{aligned} \tag{B21}$$

Recall also that the first derivatives of the Lagrangian and the Hamiltonian read,

$$L_{\mathbf{r}} = \mathbf{p} \quad , \quad L_{\mathbf{x}} = -H_{\mathbf{x}} = \dot{\mathbf{p}} \quad , \qquad L_{\mathbf{r}} \cdot \mathbf{r} = L = \mathbf{p} \cdot \mathbf{r} = v_{\text{ray}}^{-1} \quad ,$$
$$L_{\mathbf{x}} = L_{\mathbf{xr}} \mathbf{r} \quad , \quad H_{\mathbf{p}} = \dot{\mathbf{x}} = \mathbf{r} \quad , \quad \text{and} \quad L(\mathbf{x}, \mathbf{r}) = \mathbf{p} \cdot \mathbf{r} - H(\mathbf{x}, \mathbf{p}) \quad ,$$
(B22)

Both the Lagrangian $L(\mathbf{x}, \mathbf{r})$ and the Hamiltonian $H(\mathbf{x}, \mathbf{p})$ have the units of slowness.

Tangent counterpart of the Hamiltonian DRT solution

We re-emphasize that the constraint $\dot{\mathbf{w}} \cdot \mathbf{r} = 0$ is valid for the Hamiltonian $H(\mathbf{x}, \mathbf{p})$ only. Introduction of equation G29 into this constraint makes it possible to define the scalar factor $\rho_t(s)$,

$$\dot{\mathbf{w}}(s) \cdot \mathbf{r}(s) = \frac{d}{ds}\left[\mathbf{u}(s) + \rho_t(s)\mathbf{r}(s)\right] \cdot \mathbf{r}(s) = \dot{\mathbf{u}} \cdot \mathbf{r} + \dot{\rho}_t \mathbf{r} \cdot \mathbf{r} + \rho_t \dot{\mathbf{r}} \cdot \mathbf{r} = 0 \quad . \tag{B23}$$

Recall that the ray direction has a constant (unit) length, thus, $\mathbf{r}$ is normal to $\dot{\mathbf{r}}$, and the constraint yields,

$$\dot{\mathbf{u}} \cdot \mathbf{r} + \dot{\rho}_t = 0 \quad \rightarrow \quad \dot{\rho}_t = d\rho_t / ds = -\dot{\mathbf{u}} \cdot \mathbf{r} = +\mathbf{u} \cdot \dot{\mathbf{r}} \quad . \tag{B24}$$

Thus, the arclength derivative $\dot{\rho}_t(s)$ is proportional to the curvature of the central ray.

**APPENDIX C. NUMERICAL TEST FOR HESSIANS OF THE LAGRANGIAN AND HAMILTONIAN DYNAMIC FOR ANISOTROPIC MEDIA**

Triclinic medium

In order to validate the relationship in equation B15, we perform a numerical test using the same triclinic medium that was used in Appendix F of Part II, where all 21 stiffness tensor components vary in 3D space. We use the same point location and the same ray velocity direction. The following numerical values of the Hamiltonian derivatives have been obtained,

$$H_{\mathbf{x}} = -\dot{\mathbf{p}} = \begin{bmatrix} 5.4193304 \cdot 10^{-3} & 2.4547875 \cdot 10^{-2} & 2.7544272 \cdot 10^{-2} \end{bmatrix} \text{s} \cdot \text{km}^{-2} \quad , \quad \text{(C1)}$$

$$H_{\mathbf{p}} = \mathbf{r} = \begin{bmatrix} 0.224 & 0.600 & 0.768 \end{bmatrix} \quad \text{(unitless)} \quad , \quad \text{(C2)}$$

$$H_{\mathbf{xx}} = \begin{bmatrix} +3.2164007 \cdot 10^{-2} & +2.1441175 \cdot 10^{-2} & -3.4153017 \cdot 10^{-3} \\ +2.1441175 \cdot 10^{-2} & +7.4211502 \cdot 10^{-2} & +7.5019266 \cdot 10^{-2} \\ -3.4153017 \cdot 10^{-3} & +7.5019266 \cdot 10^{-2} & +1.8923660 \cdot 10^{-1} \end{bmatrix} \text{s} \cdot \text{km}^{-3} \quad , \quad \text{(C3)}$$

$$H_{\mathbf{px}} = \begin{bmatrix} +0.80382004 & +0.78194052 & +1.3854159 \\ -0.42802971 & +1.6703502 & +4.6083364 \\ +1.0412913 & +2.7309520 & +0.78011255 \end{bmatrix} \text{km}^{-1} \quad , \quad H_{\mathbf{xp}} = H_{\mathbf{px}}^{T} \quad . \quad \text{(C4)}$$

$$H_{\mathbf{pp}} = \begin{bmatrix} 31.352299 & 23.555485 & 11.361778 \\ 23.555485 & 109.22468 & 12.018191 \\ 11.361778 & 12.018191 & 120.69895 \end{bmatrix} \text{km/s} \quad , \quad \text{(C5)}$$

$$H_{\mathbf{pp}}^{-1} = \begin{bmatrix} +3.9019782 \cdot 10^{-2} & -8.0996251 \cdot 10^{-3} & -2.8665639 \cdot 10^{-3} \\ -8.0996251 \cdot 10^{-3} & +1.0938157 \cdot 10^{-2} & -3.2668655 \cdot 10^{-4} \\ -2.8665639 \cdot 10^{-3} & -3.2668655 \cdot 10^{-4} & +8.5874436 \cdot 10^{-3} \end{bmatrix} \text{s/km} \quad . \quad \text{(C6)}$$

The matrix of the directional second derivatives of the Lagrangian reads,

$$L_{rr} = \begin{bmatrix} +3.8643653 \cdot 10^{-2} & -9.1071052 \cdot 10^{-3} & -4.1561393 \cdot 10^{-3} \\ -9.1071052 \cdot 10^{-3} & +8.2395462 \cdot 10^{-3} & -3.7809068 \cdot 10^{-3} \\ -4.1561393 \cdot 10^{-3} & -3.7809068 \cdot 10^{-3} & +4.1660410 \cdot 10^{-3} \end{bmatrix} \text{ s/km} \quad . \quad \text{(C7)}$$

The eigenvalues of matrix $H_{pp}^{-1}$ are,

$$\lambda_{pp}^{inv} = \begin{bmatrix} 4.1408289 \cdot 10^{-2} & 9.6409550 \cdot 10^{-3} & 7.4961386 \cdot 10^{-3} \end{bmatrix} \text{ s/km} \quad . \quad \text{(C8)}$$

The eigenvalues of matrix $L_{rr}$ are,

$$\lambda_{rr} = \begin{bmatrix} 4.1408289 \cdot 10^{-2} & 9.6409550 \cdot 10^{-3} & 0 \end{bmatrix} \text{ s/km} \quad . \quad \text{(C9)}$$

The eigenvectors, common for matrices $H_{pp}$, $H_{pp}^{-1}$ and $L_{rr}$ are (in the columns),

$$\mathbf{V}_\lambda = \begin{bmatrix} -0.96343146 & +0.14705041 & 0.224 \\ +0.25522613 & +0.75819495 & 0.600 \\ +0.081605423 & -0.63522956 & 0.768 \end{bmatrix} . \quad \text{(C10)}$$

As we see, the last eigenvector is the ray velocity direction.

The relative error of equation B15 reads,

$$\frac{|H_x - \lambda_r H_{xp} \mathbf{r}|}{|H_x|} = 4.427 \cdot 10^{-15} \quad , \quad \text{(C11)}$$

which is the accuracy of the computer arithmetic. Equation H34 also holds within the machine precision.

Ellipsoidal orthorhombic media

We also study analytically equation B7 for an ellipsoidal orthorhombic anisotropy, with arbitrarily varying axial velocities $A_v(\mathbf{x}), B_v(\mathbf{x}), C_v(\mathbf{x})$, applying the arclength-related Hamiltonian based on its slowness surface (equation G1 of Part II). We made sure that both sides of this equation are identical functions of the location and slowness,

$$H_{\mathbf{x}} = H_{\mathbf{xp}} H_{\mathbf{pp}}^{-1} H_{\mathbf{p}} = \frac{p_1^2 A_v(\mathbf{x}) \nabla A_v(\mathbf{x}) + p_2^2 B_v(\mathbf{x}) \nabla B_v(\mathbf{x}) + p_3^2 C_v(\mathbf{x}) \nabla C_v(\mathbf{x})}{\sqrt{p_1^2 A_v^4(\mathbf{x}) + p_2^2 B_v^4(\mathbf{x}) + p_3^2 C_v^4(\mathbf{x})}} \qquad . \tag{C12}$$

For an elliptic orthorhombic medium, the eigenvalue $\lambda_{\mathbf{r}}$ of the inverse matrix $H_{\mathbf{pp}}^{-1}$, whose corresponding eigenvector is the ray velocity direction, reads,

$$\lambda_{\mathbf{r}} = -\frac{v_{\text{ray}}}{A_v^2(\mathbf{x}) r_1^2 + B_v^2(\mathbf{x}) r_2^2 + C_v^2(\mathbf{x}) r_3^2} = -\frac{v_{\text{ray}}^3}{A_v^6(\mathbf{x}) p_1^2 + B_v^6(\mathbf{x}) p_2^2 + C_v^6(\mathbf{x}) p_3^2} \qquad . \tag{C13}$$

Comment on the numerical integration of the Jacobi DRT set

In this study, we apply the variational approach to solve the Jacobi DRT equation. However, in principle, a numerical integration (e.g., with the Runge-Kutta method) is also possible. This approach requires the ODE set to be resolved for the higher derivative (wrt the arclength), in our case $\ddot{\mathbf{u}}$. By opening the brackets on the left-hand side of the Jacobi DRT equation (equation 2), and moving all terms, except that with $\ddot{\mathbf{u}}$, to the right side, we obtain,

$$L_{\mathbf{rr}} \cdot \ddot{\mathbf{u}} = \left( L_{\mathbf{xr}} - \dot{L}_{\mathbf{rr}} - L_{\mathbf{rx}} \right) \cdot \dot{\mathbf{u}} + \left( L_{\mathbf{xx}} - \dot{L}_{\mathbf{rx}} \right) \cdot \mathbf{u} \qquad . \tag{C14}$$

Indeed, since $L_{\mathbf{rr}}$ is not invertible, we cannot obtain $\ddot{\mathbf{u}}$ from this equation. This is not a surprise: only two equations of this set are independent. We note that this is the main claim in Červený's (2002a, 2002b) papers – requiring the Lagrangian to be a homogeneous function of second degree in $\mathbf{r}$ (where in our case the Lagrangian is of a first degree). Hence, in order to have a fully-defined solution one should add the normal-shift constraint, that can be differentiated twice,

$$\mathbf{u} \cdot \mathbf{r} = 0 \ , \quad \dot{\mathbf{u}} \cdot \mathbf{r} + \mathbf{u} \cdot \dot{\mathbf{r}} = 0 \ , \quad \ddot{\mathbf{u}} \cdot \mathbf{r} + 2\dot{\mathbf{u}} \cdot \dot{\mathbf{r}} + \mathbf{u} \cdot \ddot{\mathbf{r}} = 0 \qquad . \tag{C15}$$

Next, the third equation of set C15 divided by the ray velocity (in order to keep the right units), is appended to set C14. Due to the additional equation, the matrices obtain an extra line and become of dimension $4 \times 3$,

$$\begin{bmatrix} L_{\mathbf{rr}} \\ \mathbf{r}/v_{\text{ray}} \end{bmatrix} \cdot \ddot{\mathbf{u}} = \begin{bmatrix} L_{\mathbf{xr}} - \dot{L}_{\mathbf{rr}} - L_{\mathbf{rx}} \\ 2\dot{\mathbf{r}}/v_{\text{ray}} \end{bmatrix} \cdot \dot{\mathbf{u}} + \begin{bmatrix} L_{\mathbf{xx}} - \dot{L}_{\mathbf{rx}} \\ \ddot{\mathbf{r}}/v_{\text{ray}} \end{bmatrix} \cdot \mathbf{u} \qquad . \tag{C16}$$

Next, the three matrices in equation C16 are multiplied by,

$$\begin{bmatrix} L_{\mathbf{rr}} \\ \mathbf{r}/v_{\text{ray}} \end{bmatrix}^T = \begin{bmatrix} L_{\mathbf{rr}} & \mathbf{r}/v_{\text{ray}} \end{bmatrix} \qquad , \tag{C17}$$

from the left. After this operation, the dimensions of the resulting matrices in the DRT equation return to $3 \times 3$. In particular, the left side of equation I16 becomes,

$$\begin{bmatrix} L_{\mathbf{rr}} & \mathbf{r}/v_{\text{ray}} \end{bmatrix} \cdot \begin{bmatrix} L_{\mathbf{rr}} \\ \mathbf{r}/v_{\text{ray}} \end{bmatrix} = L_{\mathbf{rr}}^2 + \frac{\mathbf{r} \otimes \mathbf{r}}{v_{\text{ray}}^2} \qquad . \tag{C18}$$

The matrix on the right-hand side of equation C18 is invertible, which allows us to explicitly obtain the Jacobi set in the form, required for the numerical integration, $\ddot{\mathbf{u}} = \mathbf{A} \, \dot{\mathbf{u}} + \mathbf{B} \, \mathbf{u}$, where $\mathbf{A}$ and $\mathbf{B}$ are the corresponding matrices of dimension $3 \times 3$ with the arclength-dependent components.

Note that although only the third equation of set C15 was used in the ODE set, the initial conditions should comply with the first two equations of C15. These two equations also need to be enforced in each step of the numerical integration, to ensure computational accuracy.

We finally note that although we show that the direct integration approach is possible, we choose in this work to apply the variational approach (with the proposed finite element method) which we find very accurate; it is a natural continuation of our kinematic solution.

Comment on matrix $\mathbf{M}$

Recall that matrix $\mathbf{M}$ includes the partial second derivatives of the traveltime wrt the location components of the paraxial rays, and is defined as,

$$\mathbf{M}(s) = \frac{d^2 t}{d\mathbf{x}_{prx}^2} = \mathbf{P}(s)\mathbf{Q}^{-1}(s) \quad \text{where} \quad \begin{aligned} \mathbf{Q} &= \left[\frac{\partial \mathbf{x}_{prx}}{\partial \gamma_1} \quad \frac{\partial \mathbf{x}_{prx}}{\partial \gamma_2} \quad \frac{\partial \mathbf{x}_{prx}}{\partial s}\right] = [\mathbf{u}_1 \quad \mathbf{u}_2 \quad \mathbf{r}] \\ \mathbf{P} &= \left[\frac{\partial \mathbf{p}}{\partial \gamma_1} \quad \frac{\partial \mathbf{p}}{\partial \gamma_2} \quad \frac{\partial \mathbf{p}}{\partial s}\right] \end{aligned}. \quad (C19)$$

The first equation of set B2 makes it possible to obtain the first two columns of matrix $\mathbf{P}$, $\mathbf{p}_{\gamma_i} = \partial \mathbf{p}/\partial \gamma_i$, $i=1,2$, from the normal shifts $\mathbf{u}_i$ and their derivatives $\dot{\mathbf{u}}_i$. The third column of this matrix, $\dot{\mathbf{p}} = \partial \mathbf{p}/\partial s$, can be obtained from the kinematics, either Hamiltonian, $\dot{\mathbf{p}} = -H_{\mathbf{x}}(\mathbf{x},\mathbf{p})$, or Lagrangian, $\dot{\mathbf{p}} = L_{\mathbf{x}}(\mathbf{x},\mathbf{r})$.

We note that matrix $\mathbf{M}$ includes full derivatives: its components account for the corresponding ray direction variation when a location of a point along the ray changes. In more details, let $C$ be a running point along the central ray path between the source $S$ and the receiver $R$, where $t(S,R)$ is the traveltime between the source and receiver, and $t(S,C)$ – between the source and point $C$; in a particular case, $C$ may coincide with $R$. The location of the source is fixed. The $3 \times 3$ symmetric matrix $\mathbf{M}$ is the Hessian of the traveltime $t(S, R)$ wrt the location change of point $C$. An infinitesimal change of location $C$, $\delta \mathbf{x}_C$, is accompanied by a dependent infinitesimal change of direction $\delta \mathbf{r}_C = \delta \mathbf{r}_C (\delta \mathbf{x}_C)$ of the ray arriving to $C$. Partial derivatives in matrix $\mathbf{M}$ include this change of direction. Thus, the full spatial Hessian of the traveltime can be viewed as,

$$\mathbf{p} = \frac{dt[\mathbf{x}, \mathbf{r}(\mathbf{x})]}{d\mathbf{x}} = \frac{\partial t}{\partial \mathbf{x}} + \frac{\partial t}{\partial \mathbf{r}} \frac{d\mathbf{r}}{d\mathbf{x}} \quad , \quad \frac{d^2 t[\mathbf{x}, \mathbf{r}(\mathbf{x})]}{d\mathbf{x}^2} = \frac{d\mathbf{p}[\mathbf{x}, \mathbf{r}(\mathbf{x})]}{d\mathbf{x}} = \frac{\partial \mathbf{p}}{\partial \mathbf{x}} + \frac{\partial \mathbf{p}}{\partial \mathbf{r}} \frac{d\mathbf{r}}{d\mathbf{x}} \quad , \quad (C20)$$

where $d\mathbf{r}/d\mathbf{x}$ is an asymmetric matrix of dimension $3 \times 3$. We use notation $\mathbf{p} = dt/d\mathbf{x}$ for the slowness vector (at a current point of the ray path), where the ray direction $\mathbf{r}$ at this point varies respectively to preserve the stationary traveltime between the source and this point. With this notation, a point along the path has three independent DoF: its location components. On the other hand, for each component of $\partial t / \partial \mathbf{x}$, one location component is considered varying, while the two other location components and all three direction components are kept fixed. With this notation, the given point has six independent DoF: its location components and the ray direction components. The same convention holds for the second spatial derivatives of the traveltime. Note also that $\mathbf{p} = L_{\mathbf{r}}$, and $\partial \mathbf{p}/\partial \mathbf{x} = L_{\mathbf{rx}}$ is an asymmetric matrix (tensor), while $d\mathbf{p}/d\mathbf{x} = \mathbf{M}$ is a symmetric matrix. In equation set C20, the right-hand sides of the full first and second traveltime derivatives, wrt location components of a point along the ray, include two terms. The first term is computed assuming that the ray direction is fixed. The second term accounts for the varying ray direction at that point.

Remark: The matrices $\partial \mathbf{p}/\partial \mathbf{x}$ for a fixed ray direction $\mathbf{r}$ and $\partial \mathbf{p}/\partial \mathbf{r}$ for a fixed location $\mathbf{x}$ have been obtained analytically for general anisotropic elastic media by Ravve and Koren (2019), where obviously, in real inhomogeneous media, the computation requires numerical spatial first and second derivatives of the density-normalized stiffness tensor components.

## APPENDIX D. DISCREPANCY OF THE PLANE-WAVE SLOWNESS DIRECTION

As mentioned, in this study the term plane-wave IC refers to paraxial rays sharing the same ray direction (rather than the slowness direction) of the central ray at the source. This is suitable because in the Eigenray Lagrangian formulation, the primary DoF are the ray locations and directions. In this appendix, for a central and plane-wave paraxial rays sharing the same ray direction at the source, $\mathbf{r}_S$, we analyze the difference between the paraxial slowness vectors and the slowness vector of the central ray. The difference between the vectors depends on the derivatives of the slowness vector wrt the plane-wave RC, $\mathbf{p}_{\gamma,i,S} = \partial \mathbf{p}/\partial \gamma_{i,S}$, $i = 3, 4$, computed for the central ray. Obviously, the discrepancy between the two slowness directions depends on the strength of the anisotropy. In other words, the angles between the paraxial slowness vectors $\mathbf{p}_{\gamma,i,S}$ and the slowness vector of the central ray $\mathbf{p}_S$ at the source point are normally small for weak anisotropy and large for strong anisotropy.

Vector $\mathbf{p}_\gamma$ can be obtained from the first equation of the Hamiltonian DRT set B1,

$$\mathbf{p}_{\text{prx},\gamma,S} = H_{\mathbf{pp},S}^{-1} \left( \dot{\mathbf{w}}_S - H_{\mathbf{px},S} \mathbf{w}_S \right) \quad , \tag{D1}$$

where,

$$\mathbf{w}_S = \mathbf{u}_S + \rho_{t,S}\mathbf{r}_S \quad , \tag{D2}$$

is the Hamiltonian shift at the source point. Generally, vector $\mathbf{w}_S$ includes both, the normal and tangent counterparts, but at the source point the tangent factor $\rho_{t,S}$ vanishes. The derivative of the Hamiltonian shift vector reads,

$$\dot{\mathbf{w}}_S = \dot{\mathbf{u}}_S + \dot{\rho}_{t,S}\mathbf{r}_S + \rho_{t,S}\dot{\mathbf{r}}_S \quad . \tag{D3}$$

Equation B24 gives the derivative of the tangent factor $\dot{\rho}$, and equation D5 of Part V lists the plane-wave initial condition for the derivative $\dot{\mathbf{u}}_S$,

$$\rho_{t,S} = 0 \quad , \quad \dot{\rho}_{t,S} = \mathbf{u}_S \cdot \dot{\mathbf{r}}_S \quad , \quad \dot{\mathbf{u}}_S = -(\mathbf{u}_S \cdot \dot{\mathbf{r}}_S)\mathbf{r}_S \quad . \tag{D4}$$

Introducing equation D4 into D3, we conclude that the arclength derivative of the Hamiltonian shift vanishes at the source point of the plane wave, $\dot{\mathbf{w}}_S = 0$, while the Hamiltonian and Lagrangian shifts coincide, $\mathbf{w}_S = \mathbf{u}_S$, and equation D1 simplifies to,

$$\mathbf{p}_{\text{prx},\gamma,S} = -H_{\mathbf{pp},S}^{-1} H_{\mathbf{px},S}\, \mathbf{u}_S \quad . \tag{D5}$$

Equation set B20 relates the Lagrangian's and Hamiltonian's Hessians, in particular,

$$L_{\mathbf{rx}} = -H_{\mathbf{pp}}^{-1} H_{\mathbf{px}} \quad , \tag{D6}$$

and equation D5 reduces to,

$$\mathbf{p}_{\text{prx},\gamma_i,S} = L_{\mathbf{rx},S}\, \mathbf{u}_{i,S} \quad , \quad i = 3,4 \quad . \tag{D7}$$

A similar result can be obtained in a simpler way using the Lagrangian approach explained below. The momentum equation holds for both, the central and paraxial rays. For a a general paraxial ray, the slowness vector reads,

$$\mathbf{p}_{prx}\left(\mathbf{x}_{prx},\mathbf{r}_{prx}\right)=\mathbf{p}_{prx}\left(\mathbf{x}+\delta\mathbf{x},\mathbf{r}+\delta\mathbf{r}\right)=L_{\mathbf{r}}\left(\mathbf{x}+\delta\mathbf{x},\mathbf{r}+\delta\mathbf{r}\right) \quad . \tag{D8}$$

According to our proposed definition of a plane wave, the paraxial ray direction at the source is collinear with that of the central ray; the directional variation $\delta\mathbf{r}$ vanishes,

$$\mathbf{p}_{prx,S}\left(\mathbf{x}_S+\delta\mathbf{x}_S,\mathbf{r}_S\right)=L_{\mathbf{r}}\left(\mathbf{x}_S+\delta\mathbf{x}_S,\mathbf{r}_S\right)=L_{\mathbf{r}}\left(\mathbf{x}_S+\gamma_3\mathbf{u}_{3,S}+\gamma_4\mathbf{u}_{4,S},\mathbf{r}_S\right) \quad . \tag{D9}$$

This yields the derivative of the paraxial slowness vector,

$$\begin{aligned}\mathbf{p}_{prx,\gamma_i,S} &= \frac{\partial \mathbf{p}_{prx,S}\left(\mathbf{x}_S+\delta\mathbf{x}_S,\mathbf{r}_S\right)}{\partial \gamma_i} = \frac{\partial}{\partial \gamma_i} L_{\mathbf{r}}\left(\mathbf{x}_S+\gamma_3\mathbf{u}_{3,S}+\gamma_4\mathbf{u}_{4,S},\mathbf{r}_S\right) \\ &= \frac{\partial L_{\mathbf{r}}}{\partial \mathbf{x}}\frac{\partial \mathbf{x}_{prx,S}}{\partial \gamma_i} = L_{\mathbf{rx}}\mathbf{u}_i \quad , \quad i=3,4 \quad ,\end{aligned} \tag{D10}$$

which matches equation D7, obtained with the Hamiltonian approach.

Thus, at the source point, the slowness vector of a paraxial ray reads,

$$\mathbf{p}_{prx}\left(\gamma_3,\gamma_4,S\right)=\mathbf{p}_S+\gamma_3\,\mathbf{p}_{prx,\gamma_3,S}+\gamma_4\,\mathbf{p}_{prx,\gamma_4,S} \quad , \tag{D11}$$

where $\mathbf{p}_S$ is the slowness vector at the source point of the central ray, $\mathbf{u}_{3,S}$ and $\mathbf{u}_{4,S}$ are the normalized eigenvectors of matrix $L_{\mathbf{rr},S}$ corresponding to the nonzero eigenvalues, $\gamma_3$ and $\gamma_4$ are the plane-wave RC, and $L_{\mathbf{rx},S}$ is the mixed direction/location Hessian of the Lagrangian, that follows from equation A2 of Part V,

$$L_{\mathbf{rx},S} = -\frac{\mathbf{r} \otimes \nabla_{\mathbf{x}} v_{\text{ray}}}{v_{\text{ray}}^2} + 2\frac{\nabla_{\mathbf{r}} v_{\text{ray}} \otimes \nabla_{\mathbf{x}} v_{\text{ray}}}{v_{\text{ray}}^3} - \frac{\nabla_{\mathbf{r}} \nabla_{\mathbf{x}} v_{\text{ray}}}{v_{\text{ray}}^2}, \quad \text{at } s = S \qquad . \tag{D12}$$

Note that with the use of the auxiliary equation A26, we obtain,

$$\begin{aligned}
\partial \mathbf{p}_{\text{prx},\gamma_i,S} &\equiv \left.\frac{\partial \mathbf{p}_{\text{prx}}}{\partial \gamma_i}\right|_{s=S} = L_{\mathbf{rx},S} \mathbf{u}_{i,S} = \\
&-\frac{\mathbf{r} \otimes \nabla_{\mathbf{x}} v_{\text{ray}}}{v_{\text{ray}}^2}\mathbf{u}_{i,S} + 2\frac{\nabla_{\mathbf{r}} v_{\text{ray}} \otimes \nabla_{\mathbf{x}} v_{\text{ray}}}{v_{\text{ray}}^3}\mathbf{u}_{i,S} - \frac{\nabla_{\mathbf{r}} \nabla_{\mathbf{x}} v_{\text{ray}}}{v_{\text{ray}}^2}\mathbf{u}_{i,S} \\
&= -\frac{\nabla_{\mathbf{x}} v_{\text{ray}} \cdot \mathbf{u}_{i,S}}{v_{\text{ray}}^2}\mathbf{r} + 2\frac{\nabla_{\mathbf{x}} v_{\text{ray}} \cdot \mathbf{u}_{i,S}}{v_{\text{ray}}^3}\nabla_{\mathbf{r}} v_{\text{ray}} - \frac{\nabla_{\mathbf{r}} \nabla_{\mathbf{x}} v_{\text{ray}}}{v_{\text{ray}}^2}\mathbf{u}_{i,S} \quad .
\end{aligned} \tag{D13}$$

According to equation set A12 of Part I, a) the directional gradient of the ray velocity, $\nabla_{\mathbf{r}} v_{\text{ray}}$, is normal to the ray direction, and b) the ray direction $\mathbf{r}$ is an eigenvector of the mixed Hessian, $\nabla_{\mathbf{r}} \nabla_{\mathbf{x}} v_{\text{ray}}$, where the corresponding eigenvalue is zero. This means that the two other eigenvectors of this tensor (let's call them $\mathbf{v}_1$ and $\mathbf{v}_2$, with the eigenvalues $\mu_1$ and $\mu_2$, respectively) are in the plane normal to the ray (and are normal to each other). Vector $\mathbf{u}_{i,S}$ belongs to the normal plane as well, and can be decomposed into a linear combination of $\mathbf{v}_1$ and $\mathbf{v}_2$,

$$\mathbf{u}_{i,S} = (\mathbf{u}_{i,S} \cdot \mathbf{v}_1)\mathbf{v}_1 + (\mathbf{u}_{i,S} \cdot \mathbf{v}_2)\mathbf{v}_2 \qquad . \tag{D14}$$

and the matrix-vector product $\nabla_{\mathbf{r}} \nabla_{\mathbf{x}} v_{\text{ray}} \cdot \mathbf{u}_S$ represents a vector normal to the ray,

$$\begin{aligned}
\nabla_{\mathbf{r}} \nabla_{\mathbf{x}} v_{\text{ray}} \cdot \mathbf{u}_{i,S} &= \nabla_{\mathbf{r}} \nabla_{\mathbf{x}} v_{\text{ray}} \left[ (\mathbf{u}_{i,S} \cdot \mathbf{v}_1)\mathbf{v}_1 + (\mathbf{u}_{i,S} \cdot \mathbf{v}_2)\mathbf{v}_2 \right] \\
&= \mu_1 (\mathbf{u}_{i,S} \cdot \mathbf{v}_1)\mathbf{v}_1 + \mu_2 (\mathbf{u}_{i,S} \cdot \mathbf{v}_2)\mathbf{v}_2 \quad .
\end{aligned} \tag{D15}$$

Thus, the vector-form right-hand side of equation D13 includes an "isotropic" vector (a term with no directional Nabla), collinear to the ray direction $\mathbf{r}$, and two "anisotropic" vectors in the normal plane. In the case of weak anisotropy, the first term prevails.